\documentclass[aps,prl,twocolumn,amsmath,amssymb,superscriptaddress,longbibliography]{revtex4-1}
\usepackage[english]{babel}

\usepackage{amssymb}
\usepackage{dcolumn}
\usepackage{bm}
\usepackage{graphicx}
\usepackage{amsmath}
\usepackage{CJK}
\usepackage{graphicx}        % standard LaTeX graphics tool
\graphicspath{{pict/}{}}

\usepackage{dcolumn}%%%%%new
\usepackage{bm}
\usepackage[dvipdfm, pdfstartview=FitH, CJKbookmarks=true, bookmarksnumbered=true, bookmarksopen=true, colorlinks=true, pdfborder=001, citecolor=blue, linkcolor=blue, urlcolor=blue, linktocpage=true] {hyperref}

\begin{document}
\title{Dynamic winding number for exploring band topology}

\author{Bo Zhu}
\affiliation{Laboratory of Quantum Engineering and Quantum Metrology, School of Physics and Astronomy, Sun Yat-Sen University (Zhuhai Campus), Zhuhai 519082, China}
\affiliation{State Key Laboratory of Optoelectronic Materials and Technologies, Sun Yat-Sen University (Guangzhou Campus), Guangzhou 510275, China}

\author{Yongguan Ke}
\affiliation{Laboratory of Quantum Engineering and Quantum Metrology, School of Physics and Astronomy, Sun Yat-Sen University (Zhuhai Campus), Zhuhai 519082, China}
\affiliation{Nonlinear Physics Centre, Research School of Physics, The Australian National University, Canberra ACT 2601, Australia}

\author{Honghua Zhong}
\affiliation{Institute of Mathematics and Physics, Central South University of Forestry and Technology, Changsha 410004, China}

\author{Chaohong Lee}
\altaffiliation{lichaoh2@mail.sysu.edu.cn.}
\affiliation{Laboratory of Quantum Engineering and Quantum Metrology, School of Physics and Astronomy, Sun Yat-Sen University (Zhuhai Campus), Zhuhai 519082, China}
\affiliation{State Key Laboratory of Optoelectronic Materials and Technologies, Sun Yat-Sen University (Guangzhou Campus), Guangzhou 510275, China}
\affiliation{Nonlinear Physics Centre, Research School of Physics, The Australian National University, Canberra ACT 2601, Australia}

\begin{abstract}
Topological invariants play a key role in the characterization of topological states.
Due to the existence of exceptional points, it is a great challenge to detect topological invariants in non-Hermitian systems.
We put forward a dynamic winding number, the winding of realistic observables in long-time average, for exploring band topology in both Hermitian and non-Hermitian two-band models via a unified approach.
We build a concrete relation between dynamic winding numbers and conventional topological invariants.
In one-dimension, the dynamical winding number directly gives the conventional winding number.
In two-dimension, the Chern number relates to the weighted sum of dynamic winding numbers of all phase singularity points.
This work opens a new avenue to measure topological invariants not requesting any prior knowledge of system topology via time-averaged spin textures.
\end{abstract}
\date{\today}

\maketitle

{\it Introduction}. Topological invariant, a global quantity defined with static Bloch functions, has been widely used for classifying and characterizing topological states of matters, including insulators, superconductors, semimetals and waveguides etc~\cite{qi2011topological, bernevig2013topological, ando2013topological, chiu2016classification, lv2017observation, ozawa2019topological}.
Nontrivial topological invariants attribute to novel topological effects, such as winding number for quantized geometric phase~\cite{cao2018unifying, yin2018geometrical}, and Chern number for integer Hall effect~\cite{thouless1982quantized, hatsugai1993chern} and Thouless pumping~\cite{thouless1983quantization, ke2016topological, lohse2018exploring}.
Measuring topological invariants provide undoubted evidence of topological states, beneficial for precision measurement~\cite{klitzing1980new, cooper2015adiabatic}, error-resistant spintronics~\cite{nayak2008non, pesin2012spintronics} and quantum computing~\cite{haldane2017nobel, lian2018topological}.

Most existed methods to measure topological invariants are based on adiabatic band sweeping~\cite{atala2013direct, jotzu2014experimental, aidelsburger2015measuring, flaschner2016experimental,hu2019dispersion}. %and quantum walks~\cite{cardano2017detection, wang2019direct}.
However, these methods do not work well for imperfect initial states and small energy gaps and become invalid for non-Hermitian systems.
In recent, topological invariants have been measured via linking numbers and band-inversion surfaces in quench dynamics~\cite{wang2017scheme, qiu2018fixed, sun2018uncover,sun2018uncover, zhang2018dynamical, tarnowski2019measuring, zhang2019dynamical}.
However, these quench schemes request prior knowledge of topology before and after quench.

As non-Hermitian systems may exhibit complex spectra and exceptional points (EPs)~\cite{heiss2012physics, hu2017exceptional, hassan2017dynamically}, their topological states have stimulated extensive interests~\cite{esaki2011edge, liang2013topological, malzard2015topologically, lee2016anomalous, leykam2017edge, rakovszky2017detecting, lieu2018topological, alvarez2018non, yoshida2018non, zhou2018non, chen2018hall,  yoshida2019symmetry, Dan2019Modes, Tsuneya2019Non}.% hu2017exceptional,bender2007making,xu2017weyl,menke2017topological,gonzalez2017topological,
Conventional topological invariants such as winding number and Chern number have been generalized to non-Hermitian systems~\cite{lee2016anomalous,yin2018geometrical,shen2018topological}, and new topological invariants such as vorticity have been introduced~\cite{Ghatak_2019}.
Due to the EPs,  the winding number in non-Hermitian systems may take half-integers~\cite{lee2016anomalous,yin2018geometrical, jiang2018topological,jin2019bulk}.
Besides, non-Bloch definition of Chern number strictly gives the numbers of chiral edge modes~\cite{yao2018edge, yao2018non, Ghatak2019Observation}.
How to measure these topological invariants is more challenging than that in Hermitian systems.
For an example, the Hall conductivity is no longer quantized despite being classified as a Chern insulator based on non-Hermitian topological band theory~\cite{philip2018loss, chen2018hall}.
In one-dimension, the winding number in a non-Hermitian system has been determined via the mean displacement in long-time quantum walk~\cite{rudner2009topological, zeuner2015observation}, but it does not works for measuring Chern numbers and half-integer winding numbers.
One may ask, \emph{is there a unified dynamic approach for measuring topological invariants in both Hermitian and non-Hermitian systems?}

In this Letter, we study a generic two-band model which supports nontrivial topological invariants in both Hermitian and non-Hermitian regions.
We define a dynamic winding number (DWN) for the time-averaged spin textures, which is robust against various initial states.
In one-dimension, we prove that the DWNs directly gives the conventional winding numbers in both chiral- and non-chiral-symmetric systems.
In two-dimension, the Chern number relates to the weighted sum of DWNs around all singularity points (SPs), where the weight is $+1$ for the north pole and $-1$ for the south pole.
When the system change from Hermitian to non-Hermitian, each singularity point will split into two EPs (which are also SPs), the Chern number can still be extracted via the DWNs of all EPs.
Without requesting any prior knowledge of their topology, our approach provides a general guidance for measuring topological invariants in both Hermitian and non-Hermitian systems.

 {\it Dynamic winding number}. We consider a general two-band model in $d-$dimension.
The Hamiltonian in momentum space is composed of three Pauli matrices,
\begin{eqnarray} \label{Ham}
H(\bm{k})=h_x(\bm{k})\sigma_x+h_y(\bm{k})\sigma_y+h_z(\bm{k})\sigma_z.
\end{eqnarray}
Here, $\bm{k}$ is the quasi-momentum, $h_{x(y,z)}$ are periodic functions of $\bm{k}$.
The Hamiltonian could be Hermitian $H^\dag=H$ or non-Hermitian $H^\dag\ne H$.
Then, the right and left eigenvectors are respectively given by $H(\bm{k})|\varphi_{\mu}\rangle=\varepsilon_{{\mu}}|\varphi_{\mu}\rangle$ $( H^\dagger(\bm{k})|\chi_{\mu}\rangle=\varepsilon^*_{{\mu}}|\chi_{\mu}\rangle)$, where $\mu=\pm$, and $\varepsilon_{{\pm}}=\pm(h_x^2+h_y^2+h_z^2)^{1/2}$ are the eigenvalues.
For Hermitian systems, $|\varphi_{\mu}\rangle=|\chi_{\mu}\rangle$ and $\varepsilon_{\mu}=\varepsilon_{\mu}^*$.
For non-Hermitian systems, neither the eigenstates $|\varphi_{\mu}\rangle$ nor $\langle\chi_{\mu}|$ are orthogonal.
We adopt biorthogonal vectors which fulfil $\langle\chi_{\nu}| \varphi_{\mu}\rangle=\delta_{\nu,\mu}$ and $\sum_{{\mu}}|\varphi_{\mu} \rangle \langle \chi_{\mu} |=1$ by normalizing  $|\varphi_{\mu}\rangle=|\varphi_{\mu}\rangle/N_{\mu}$ and $\langle\chi_{\mu}|=\langle\chi_{\mu}|/N_{\mu}$ with $N_{\mu}=\sqrt{{\langle\chi_{\mu}| \varphi_{\mu}\rangle}}$.

For an arbitrary initial state $|\psi_{\bm{k}}(0)\rangle=\sum_{\mu} c_{{\mu}}|\varphi_{\mu}\rangle$ and its associated state
$\langle\tilde{\psi}_{\bm{k}}(0)|=\sum_{\mu} c_{{\mu}}^{*}\langle\chi_{\mu}|$,
the time-evolution of $|\psi_{\bm{k}}(t)\rangle$ and $\langle\tilde{\psi}_{\bm{k}}(t)|$ respectively satisfy $|\psi_{\bm{k}}(t)\rangle=\sum_{\mu} c_{\mu} e^{-i\varepsilon_{{\mu}} t}|\varphi_{\mu}\rangle$ and $\langle\tilde{\psi}_{\bm{k}}(t)|=\sum_{\mu} c_{{\mu}}^{*} e^{i\varepsilon_{\mu}^{*} t}\langle\chi_{\mu}|$.
According to the biorthogonal quantum mechanics~\cite{brody2013biorthogonal}, the spin
textures are given by the expectation values of Pauli matrices, $\langle \sigma_j (\bm{k},t)\rangle ={\langle \tilde{\psi}_{\bm{k}}(t) | \sigma_j |\psi_{\bm{k}}(t)\rangle}/{\langle \tilde{\psi}_{\bm{k}}(t) | \psi_{\bm{k}}(t)\rangle}$,
where $j\in {x,y,z}$.
We are interesting in its long-time average,
%\begin{eqnarray}
$\overline{\sigma_j}(\bm k)=\lim_{T\rightarrow \infty} \frac{1}{T}\int_0^{T} \langle \sigma_j (\bm{k},t)\rangle dt$.
%\end{eqnarray}
%
As the quasimomentum continuously varies, $(\overline{\sigma_i}, \overline{\sigma_j})$ will form a trajectory in the polarization plane.
The DWN of the spin vector $(\overline{\sigma_i}, \overline{\sigma_j})$ is defined as
\begin{eqnarray}
  w_d=\frac{1}{2\pi}\oint_{S} \partial_{\bm k}\eta_{ji}(\bm k) d\bm{k}, \label{dynwinding}
\end{eqnarray}
where $S$ is a close loop in parameter space $\bm{k}$, and dynamical azimuthal angle $\eta_{ji}(\bm k)= \arctan [{\overline{\sigma_j}(\bm k)}/{\overline{\sigma_i}(\bm k)}]$.
It is easy to prove that the DWN is convergent to the equilibrium azimuthal angle,
 \begin{eqnarray} \label{eta}
   \eta_{ji}=\phi_{ji}\equiv\arctan\left[h_j(\bm{k})/h_i(\bm{k})\right], \label{dynwinding2}
 \end{eqnarray}
if $|c_{+}|^2\ne|c_{-}|^2$ for Hermitian systems and $|c_{+}|^2\ne 0\bigcap |c_{-}|^2\ne 0$ for non-Hermitian systems~\cite{Suppl}.
For Hermitian systems, the DWN can be directly probed by the long-time average of spin textures.
For non-Hermitian systems, $\eta_{ji}$ is a complex angle which cannot be directly observed.
This problem can be fixed by decomposing the azimuthal angle into real and imaginary parts.
We find that only the real part of $\eta_{ji}$ contributes to the DWN and it satisfies,
\begin{eqnarray}
\Re(\eta_{ji})=\frac{1}{2}(\phi_{ji}^{RR}+ \phi_{ji}^{LL})+n \frac{\pi}{2},
\label{phiab}
\end{eqnarray}
 where $\phi_{ji}^{RR}=\arctan\big({\overline{\langle\psi_{\bm{k}} |\sigma_j|\psi_{\bm{k}}\rangle}}/{\overline{\langle\psi_{\bm{k}} |\sigma_i|\psi_{\bm{k}}\rangle}}\big)$ and $\phi_{ji}^{LL}=\arctan\big({\overline{\langle\tilde{\psi}_{\bm{k}} |\sigma_j|\tilde{\psi}_{\bm{k}}\rangle}}/{\overline{\langle\tilde{\psi}_{\bm{k}} |\sigma_i|\tilde{\psi}_{\bm{k}}\rangle}}\big)$ are both real~\cite{Suppl}, $\Re(\eta_{ji})$ represents the real part of $\eta_{ji}$.
Thus we have
%\begin{eqnarray} \label{equaion_D_ab}
$w_d=\frac{1}{2}(w_d^{RR}+w_d^{LL})$,
%\end{eqnarray}
where $w_d^{\tau}=\frac{1}{2\pi}\oint_S \partial_{\bm{k}}\phi_{jl}^{\tau} d\bm{k}$, $\tau\in RR, LL$.
This means that the DWN can also be observed by the time-evolution of left-left and right-right spin textures whose dynamics are respectively governed by $\hat H$ and $\hat H^\dag$.
In the following, we show how to utilize the DWN to uncover the topology in both Hermitian and non-Hermitian systems.

 {\it Connection between conventional winding number and dynamic winding number}. In one-dimension, if $h_z=0$, the Hamiltonian~\eqref{Ham} has chiral symmetry $\Gamma H(k) \Gamma =-H(k)$ with $\Gamma=i \sigma_x \sigma_y$ and $\bm k\rightarrow k$.
The conventional winding number $w_{\pm}$ for the Hamiltonian~\eqref{Ham} reads as,
\begin{eqnarray} \label{equaion1b}
%w_{\pm}=\frac{1}{2\pi}\oint_c dk \frac{h_x\partial_k h_y-h_y\partial_k h_x}{h_x^2+h_y^2}=\frac{1}{2\pi}\oint_c \partial_k \phi_{yx} dk,\nonumber
w_{\pm}=\frac{1}{2\pi}\oint_S dk \frac{h_x\partial_k h_y-h_y\partial_k h_x}{(\varepsilon_{{\pm}})^2},
\label{phiabded}
\end{eqnarray}
which associates with the Zak phase.
According to Eqs.~\eqref{dynwinding} and~\eqref{phiabded}, one can find that $w_{\pm} = w_d$.
%
%
%where the azimuthal angle $\phi_{yx}=\arctan(h_y/h_x)$.
%%
%According to Eq.~\eqref{eta}, we have $w_{\pm}=w_d$ and thus the sum of two conventional winding number is equal to twice as the DWN,
%\begin{eqnarray}
%w_{t}=w_{+}+w_{-}=2w_d.
%\end{eqnarray}

If $h_z\ne 0$, the Hamiltonian~\eqref{Ham} breaks the chiral symmetry.
The winding numbers $w_{\pm}$ can be given as,
\begin{eqnarray} \label{equaion1b}
w_{\pm}=\frac{1}{2\pi}\oint_S dk \frac{h_x\partial_k h_y-h_y\partial_k h_x}{\varepsilon_{{\pm}}(\varepsilon_{{\pm}}-h_z)}.
\label{phiabded2}
\end{eqnarray}
Unlike the systems with chiral symmetry, the conventional winding number for each band is not a quantized number, which indicates that $w_{\pm}$ is no longer a topological invariant.
However, the sum of two conventional winding numbers,
\begin{eqnarray} \label{equaion1d}
w_{t}=w_{+}+w_{-}=\frac{1}{\pi}\oint_S dk \frac{h_x\partial_k h_y-h_y\partial_k h_x}{h_x^2+h_y^2},
\end{eqnarray}
relates to the dynamic winding number via $w_t=2w_d$ and thus it can be used as a topological invariant.

%%%%%%%%%%%%%%%%%%%%%%%%%%%%%%%%
\begin{figure}[htp]
\center
\includegraphics[width=3.4in]{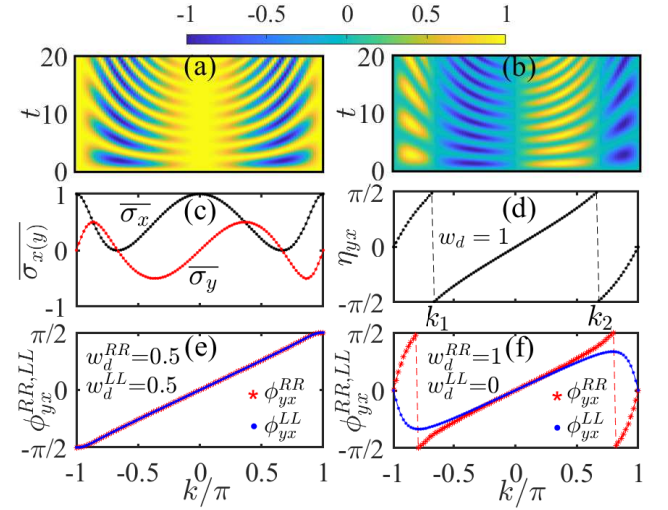}
  \caption{Extract conventional winding number via dynamic winding number. Hermitian case: (a) and (b) respectively show the time-evolution of the spin textures $\langle\sigma_x\rangle$ and $\langle\sigma_y \rangle$,
(c) time-averaged spin textures $ \overline{\sigma_x}$ (black line) and $ \overline{\sigma_y}$ (red line) as a function of $k$,
and (d) the dynamical azimuthal angle $\eta_{yx}$ as a function of $k$.
In which, $k_1$ and $k_2$ are discontinuity points.
Non-Hermitian case: $\phi_{yx}^{RR}$ and $\phi_{yx}^{LL}$ as a function of $k$ for (e) chiral-symmetric system with $h_z=0$ and (f) non-chiral-symmetric system with $h_z=0.5$.}
\label{fig2}
\end{figure}
%%%%%%%%%%%%%%%%%%%%%%%%%%%%%

%As an example, we consider a chiral-symmetric system with
As an example, we consider a system with $h_x=J_0+J_1\cos(k)$, $h_y=J_1\sin(k)-i\delta$ and $h_z=0$.
In the Hermitian case, the parameters are chosen as $\delta=0$ and $J_1=1$.
%
%As $J_0$ varies,
The conventional winding number $w_{\pm}=1$ for $|J_0|<J_1$, and $w_{\pm}=0$ for $|J_0|>J_1$.
We first calculate the time-evolution of $\langle \sigma_{x(y)}(k,t)\rangle$ and their long-time average with $J_0=0.5J_1$, see  Figs.~\ref{fig2}(a)-(c).
The spin textures $\langle\sigma_{x(y)}(k,t)\rangle$ oscillate with a momentum-dependent period $\widetilde{t}_k=\pi/|\varepsilon_{\mu}|$, and their long-time averages $\overline{\sigma_{x(y)}}$ depend on quasi-momentum, see the black and red lines in Fig.~\ref{fig2} (c).
With $\overline{\sigma_x}$ and $\overline{\sigma_y}$, we calculate $\eta_{yx}$ as a function of $k$ in Fig.~\ref{fig2}(d), where two discontinuity points $k_1$ and $k_2$ appear.
The DWN can be obtained via the integral of piecewise function,
\begin{eqnarray} \label{equaion10a}
w_d&=&\frac{1}{2\pi}\big(\int_{-\pi}^{k_1} \partial_k\eta_{yx} dk+\int_{k_1}^{k_2} \partial_k\eta_{yx} dk+\int_{k_2}^\pi \partial_k\eta_{yx} dk\big).\nonumber
\end{eqnarray}
We find that the DWN is equal to $1$, the same as the conventional winding number $w_{\pm}$.

When $\delta\ne 0$, the system becomes non-Hermitian and one always needs to measure both $\phi_{ji}^{RR}$ and $\phi_{ji}^{LL}$ to extract the DWN.
For a chiral-symmetric system (whose parameters are given as $J_0=J_1=1$, $\delta=0.3$, $h_z=0$ and $w_{\pm}=1/2$), the two dynamic azimuthal angles $\phi_{ji}^{RR}=\phi_{ji}^{LL}$, and $w_d^{RR}=w_d^{LL}=\frac{1}{2}$, see in Fig.~\ref{fig2}(e).
It means that we only need to measure $\phi_{ji}^{RR}$ or $\phi_{ji}^{LL}$ in experiments.
For a non-chiral-symmetric system (whose parameters are given as $J_0=J_1=1$, $\delta=0.3$, $h_z=0.5$, and $w_{t}=1$), we find $\phi_{ji}^{RR}\neq\phi_{ji}^{LL}$, and $w_d^{RR}=1, w_d^{LL}=0$, see Fig.~\ref{fig2}(f).
%
%We find that $w_d^a=1$ and $w_d^b=0$.
%
Nevertheless, the conventional winding number can be obtain by measuring the $w_d=(w_d^{RR}+w_d^{LL})/2$ in both chiral and non-chiral symmetric systems.

{\it Connection between Chern number and dynamic winding number}.
By generalizing the concept of gapped band structures from Hermitian to non-Hermitian systems,
the Chern number for an energy separable band can be constructed in a similar way~\cite{shen2018topological}.
In contrast to Hermitian systems, there are left-right, right-right, left-left, right-left Chern numbers in non-Hermitian systems, dependent on the definitions of Berry connection, $A_{\bm{k}}^{LR}= i \langle \chi_\mu| \partial_{\bm{k}} |\varphi_\mu\rangle$, $A_{\bm{k}}^{RR}= i \langle \varphi_\mu| \partial_{\bm{k}} |\varphi_\mu\rangle$, $A_{\bm{k}}^{LL}= i \langle \chi_\mu| \partial_{\bm{k}} |\chi_\mu\rangle$, and $A_{\bm{k}}^{RL}= i \langle \varphi_\mu| \partial_{\bm{k}} |\chi_\mu\rangle$.
Although the corresponding Berry curvatures are locally different quantities, but the four kinds of Chern numbers are the same~\cite{shen2018topological}.
Here, we only focus on analyzing the Chern number defined with left-right Berry connection $A_{\bm{k}}^{LR}=i \langle \chi_\mu| \partial_{\bm{k}}|\varphi_\mu\rangle$, which naturely reduces the Chern number in Hermitian systems as $|\varphi_{\mu}\rangle=|\chi_{\mu}\rangle$.

\begin{figure}[htp]
\center
\includegraphics[width=3.4in]{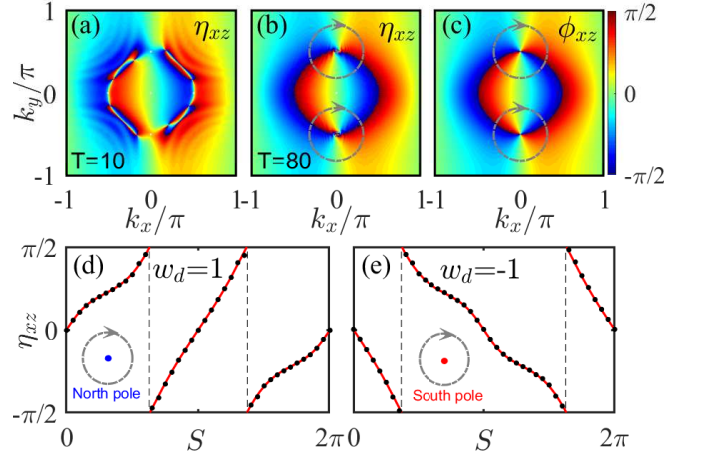}
\caption{Topologically nontrivial phase with Chern number $C=1$.
(a) and (b) respectively show $\eta_{xz}(k_x,k_y)$ obtained in the evolved time $T=10$ and $T=80$, and (c) displays $\phi_{xz}(k_x,k_y)$.
(d) and (e) show how $\eta_{xz}$ (black dots)  and $\phi_{xz}$ (red line)  change along the trajectory around the north and south poles in (b) and (c), respectively.
}\label{Fig2Chern}
\end{figure}

We map the Hamiltonian to a normalized vector,
$\vec{n}(\bm{k})=(\sin(\theta_i)\cos(\phi_{jl}), \sin(\theta_i)\sin(\phi_{jl}), \cos(\theta_i))$, which reduces a Bloch vector in Hermitian systems.
Here, $\theta_i$ denotes the angle between the vector and the axis-$i$, and $\phi_{jl}$ denotes the equilibrium azimuthal angle in the $j-l$ plane.
The reference axis is free to choose without affecting the validity of the dynamic approach.
%
%Due to $\hat H\ne \hat H^\dag$, the normalized Bloch vector satisfies $\vec{n}\ne \vec{n}^*$.
%
Then, the left and right eigenstates for the low-energy band are given as
\begin{eqnarray} \label{equaion10sa_2019}
&&\langle\chi_-(\theta_i,\phi_{jl})|=\begin{pmatrix}-e^{i\phi_{jl}/2}\cos(\frac{\theta_i}{2}),
e^{-i\phi_{jl}/2}\sin(\frac{\theta_i}{2})
\end{pmatrix}, \nonumber \\
&&|\varphi_-(\theta_i,\phi_{jl})\rangle=\begin{pmatrix}-e^{-i\phi_{jl}/2}\cos(\frac{\theta_i}{2})\\
e^{i\phi_{jl}/2}\sin(\frac{\theta_i}{2})
\end{pmatrix}.
\end{eqnarray}
%which depend respectively on the virtual normalized vector field $\vec{n}(\bm{k})=\vec{h}(\bm{k})/|\vec{h}(\bm{k})|$ and $\vec{n}^\dag(\bm{k})$.
%
% where $\theta$ and $\phi$ are generally complex.
The right and left eigenstates have a phase singularity at $\vec{n}(\bm{k})=(0, 0,\pm 1)$, in which $+$ and $-$ respectively correspond to north and south poles.
In the parameter space $(k_x,k_y)$, the location $\bm{k}_0$ of the poles satisfy $h_j(\bm{k}_0)^2+h_l(\bm{k}_0)^2=0$.
%
%The north and south EPS correspond to $\theta=0$  and $\pi$, respectively.
%
The left-right Berry connection ${A}_{k_{x(y)}}^{LR}$  of the low-energy band are given by
\begin{eqnarray} \label{equaion10sa19}
{A}_{k_{x(y)}}^{LR}=i\langle \chi_-|\partial_{k_{x(y)}}|\varphi_-\rangle=\frac{\cos(\theta_i)}{2}\frac{\partial\phi_{jl}}{\partial_{k_{x(y)}}}. \nonumber
\end{eqnarray}
We discreterize the parameter space $(k_x,k_y)$ by $N\times M$ mesh grids in the first Briliouin zone~\cite{fukui2005chern}.
For each grid, a direct application of the two-dimensional Stokes theorem implies~\cite{asboth2016short}
\begin{eqnarray} \label{equaion10sa13}
C^{LR}=\frac{1}{2\pi}\sum\limits_{l_x=1}^N\sum\limits_{l_y=1}^M\oint_{S_{l_x,l_y}}(A_{k_x}dk_x+A_{k_y}dk_y),
\end{eqnarray}
where $S_{l_x,l_y}$ represents the clockwise path integration of the $(l_x,l_y)$ grid.
We find that the Chern number is determined by the winding numbers for all SPs, where $\cos(\theta_i)=h_i(\bm{k_0})/|\vec{h}(\bm{k_0})|=sgn(\Re[h_i(\bm{k_0})])=1$ for the north SPs and $\cos(\theta_i)=-1$ for the south SPs,  $\Re[h_i(\bm{k_0})]$ represents the real part of $h_i(\bm{k_0})$, and $\vec{h}=(h_x, h_y, h_z)$.
% with $h_i^R=Re(h_i)$.
%
At last, we can deduce the left-right Chern number as
\begin{eqnarray} \label{Chern}
%C^{LR}&=&\frac{1}{4\pi}\sum\limits_{\bm{k_0} \in poles} sgn(h_i^R(\bm{k_0}))\oint_{\bm{k_0}} \partial_{\bm{k}}\phi_{jl} d\bm{k} \nonumber \\
C^{LR}=\frac{1}{2}\sum\limits_{\bm{k}_0 \in \rm{SPs}} sgn(\Re[h_i(\bm{k_0})])w_d(\bm{k}_0),
\end{eqnarray}
where $w_d(\bm{k}_0)$ is the DWN for the SP at ${\bm k}_0$~\cite{Suppl}.
In non-Hermitian case, $w_d(\bm{k}_0)$ is relevant to two real angles $\phi_{jl}^{RR}$ and $\phi_{jl}^{LL}$, which can be respectively extracted via right-right spin textures $\langle {\psi}_{\bm{k}}(t)| \sigma_{j(l)}| \psi_{\bm{k}}(t)\rangle$ and left-left spin textures $\langle\tilde{\psi}_{\bm{k}}(t)| \sigma_{j(l)}| \tilde{\psi}_{\bm{k}}(t)\rangle$.

As an example, we consider
%a two-band model which is characterized by a unconstrained vector $\vec{h}(\bm{k})=(h_x, h_y, h_z)$ with
$h_x=J_x \sin(k_x)$, $h_y=J_y \sin(k_y)$ and $h_z=m_z-J_z \cos(k_x)-J_z \cos(k_y)-i \delta$~\cite{Suppl}.
Here, $J_{x(y,z)}$ denote spin-orbit coupling parameters, $m_z$ is the effective magnetization, and $\delta$ is a gain or loss strength.
When $\delta=0$, the system is a quantum anomalous Hall model~\cite{liu2014realization}, which has  been realized in recent experiments~\cite{chang2013experimental, sun2018uncover}.
In the Hermitian case ($J_{x(y,z)}=1,\ m_z=1,\  \delta=0$), the north and south poles in the parameter space $(k_x,k_y)$ can be determined as following.
Since the poles are related to the chosen axis, we select $\theta=\theta_y=\arccos(h_y/|\vec{h}(\bm{k})|)$ and $\phi=\phi_{xz}=\arctan(h_x/h_z)$.
However, the validity of our dynamic approach is independent on the choice of reference axis~\cite{Suppl}.
In the parameter space $(k_x,k_y)$, by solving $h_x^2+h_z^2=0$, we find that the north and south poles locate at $\bm{k}_0=(k_x,k_y)=(0,\pm \pi/2)$.% which respectively correspond to the blue and red dots in Fig.~\ref{Fig2Chern} (a).
%
%Here, $sgn[h_y(\pm \pi/2)]=\pm 1$ respectively correspond to the north and  south poles.
%
Then, we need to extract the DWN around the two poles.
We randomly choose an initial state $|\psi_{\bm{k}}(0)\rangle=\sum_{\mu} c_{{\mu}}|\varphi_{\mu}\rangle$ with $|c_{+}|^2>|c_{-}|^2$ and calculate the spin textures $\left\langle \sigma_x(\bm k, t)\right\rangle $ and $\left\langle \sigma_z(\bm k, t)\right\rangle$.
%in the time-evolution governed by $i \partial_t |\psi(t)\rangle=H (\bm k)|\psi(t)\rangle$.
%
%Here, the total evolution time are set as $T=80$.
%
The dynamical azimuthal angle $\eta_{xz}(\bm k)$ can be extracted via long-time average values $\overline{\sigma_{x(z)}}(\bm k)$, see Figs.~\ref{Fig2Chern}(a) and~\ref{Fig2Chern}(b).
We also calculate the equilibrium azimuthal angle $\phi_{xz}(\bm k)$ via the eigenstates, see Fig.~\ref{Fig2Chern}(c).
The difference between $\eta_{xz}(\bm k)$ and $\phi_{xz}(\bm k)$ gradually disappears with the increase of total time $T$.
%
%The dynamical azimuthal angle $ \eta_{xz}(\bm k)$ and equilibrium azimuthal angle $\phi_{xz}(\bm k)$ are almost the same, that is, $\eta_{xz}(\bm k)$ converges to $ \phi_{xz}(\bm k)$ in long time.
%%enclosing a single singularity point
Thus we can obtain the DWNs for the north and south poles via integrating the gradient of $\eta_{xz}(\bm k)$ in Fig.~\ref{Fig2Chern}(d) and~\ref{Fig2Chern}(e), respectively.
The DWNs for the north and south poles are respectively given as $w_d=\pm 1$.
Applying Eq.~\eqref{Chern}, one can obtain the Chern number as $1$, which is consistent with the one calculated via integrating the static Berry curvature over the whole in the parameter space $(k_x,k_y)$.

\begin{figure}[htp]
\center
\includegraphics[width=3.5in]{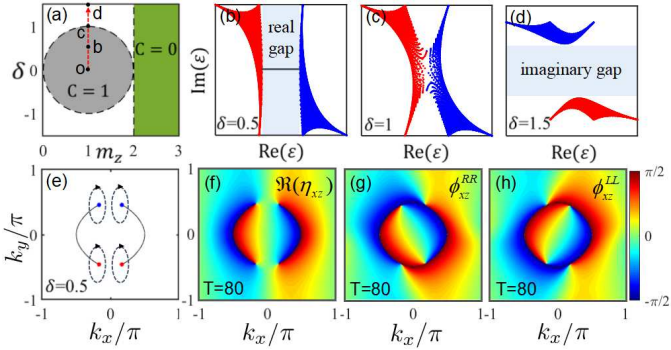}
\caption{(a) Topological phase diagram.
% the white region does not support real gap and there is no well-defined Chern number, the gray and green regions supports real gap and the Chern number $C=1$ and $0$, respectively.
%  The gray region supports real gap and the Chern number $C=1$. The green region supports real gap and the Chern number $C=0$.
  (b)-(d) Energies of bulk bands (red and blue regions) and edge mode (black line) in the complex-energy plane, corresponding to the parameter points $b$-$d$ in (a).
  (e) North EPs (blue dots) and south EPs (red dots) with $\delta=0.5$.
  (f)-(h) Dynamic azimuthal angles $\Re[\eta_{xz}(k_x,k_y)]$, $\phi_{xz}^{RR}(k_x,k_y)$ and $\phi_{xz}^{LL}(k_x,k_y)$, which are defined with left-right, right-right and left-left spin textures in the evolved time $T=80$.
}\label{fig_A2D_xin_20190227}
\end{figure}

In the more generalized case, we first show the topological phase diagram in the parameter plane ($m_z,\delta$) by setting $J_{x(y,z)}=1$, see Fig.~\ref{fig_A2D_xin_20190227}(a).
%
%The Chern numbers of the first band are $C=0$ in the green region, $C=1$ in the grey region, and not well defined in the white region.
The Chern numbers of the first band are $C=0$ and $1$ in the green and grey regions, and not well defined in the white region.
The boundaries satisfy $(m_z-1)^2+\delta^2<1$ for the gray region and $m_z>2$ for the green region.
Varying $\delta$ along the dashed red arrow, in Figs.~\ref{fig_A2D_xin_20190227}(b)--(d) we explore the correspondence between the Chern number and energy modes under open boundary condition, corresponding to the parameter points $b$-$d$.
For the topological nontrivial phase, one can see that the complex bulk bands are still gapped and the edge-state modes are still preserved in the real energy axis, see Fig.~\ref{fig_A2D_xin_20190227}(b).
Now we consider $\delta=0.5$ and keep other parameters the same as those in Fig.~\ref{Fig2Chern}(b).
We find that the two poles in Hermitian case are split into four EPs, see Fig.~\ref{fig_A2D_xin_20190227}(e).
%%
%we can obtain the left-right Chern number via similar procedure.
%%
%As $\delta$ increases, the two poles in Hermitian case are split into four EPs, see Fig.~\ref{Fig2Chern}(d).
%%
The blue and red points represent north and  south EPs, respectively.
%
%Similarly, the north EPs have $sgn(h_y^R)=1$ and the south ones have $sgn(h_y^R)=-1$.
%
In Figs.~\ref{fig_A2D_xin_20190227}(f), we give $\Re(\eta_{xz})$ in parameter space $(k_x,k_y)$, and the DWNs can be obtain by
integral of the dynamic azimuthal-angle gradient along the trajectory enclosing the north and south EPs.
Here, the DWNs for the north and south EPs are $w_d=1/2$ and $-1/2$, respectively.
Although the SPs are doubled, as each DWN is reduced by half, the Chern number keep unchanged.
For completeness, we also show the dynamic azimuthal angles defined with the right-right and left-left spin textures, see Fig.~\ref{fig_A2D_xin_20190227}(g) and~\ref{fig_A2D_xin_20190227}(h) .
The dynamic azimuthal angles are quite different from each other, corresponding to $\phi_{xz}^{RR}$ and $\phi_{xz}^{LL}$, respectively.
Nevertheless, the north and south EPs are the same as those in the Fig.~\ref{fig_A2D_xin_20190227}(e).
Around an EP, one can extract the right-right and left-left DWN $w_d^{RR}$ and $w_d^{LL}$, which satisfy $w_d=\frac{1}{2}(w_d^{RR}+w_d^{LL})$.
%
%Contrast to Fig.~\ref{fig_A2D_xin_20190227} (b) and (c), one can clearly see the dynamic winding number $w_d^a=w_d^b$ for .
%%
One important thing is that the $\phi_{xz}^{RR}$ and $\phi_{xz}^{LL}$ defined with the real left-left and right-right spin textures are accessible in experimental measurements.

In addition, our approach can also extract larger Chern numbers without extra efforts~\cite{Suppl}, which is very hard to access by adiabatic band sweeping.

{\it Conclusions and Discussions.} We put forward a new concept of dynamic winding number (DWN) and uncover its connection to conventional topological invariants in both Hermitian and non-Hermitian models.
Given a time-averaged spin texture in the parameter space, a DWN is given by a loop integral of the dynamic azimuthal-angle gradient enclosing a single singularity point.
We find that, (i) the conventional winding numbers in one-dimensional systems can be directly given by the corresponding DWNs, and (ii) the Chern numbers in two-dimensional systems relates to the weighted sum of all corresponding DWNs.
Our scheme has two main advantages.
Firstly, in contrast to the quench schemes via measuring linking numbers~\cite{wang2017scheme, tarnowski2019measuring} and band-inversion surfaces~\cite{zhang2018dynamical, sun2018uncover}, which request prior knowledge of topology before and after quench, our scheme does not request any prior knowledge.
Secondly, our scheme can be used to measure half-integer winding numbers in non-Hermitian one-dimensional systems, which can not be measured via previous methods.

Our scheme is readily realized in various systems, ranging from cold atoms in optical lattices, optical waveguide arrays, to optomechanical devices.
%~\cite{sun2018highly}~\cite{zeuner2015observation}~\cite{xu2016topological}
%
In Supplementary Material~\cite{Suppl}, we provide more details about the experimental realization of our scheme via cold atoms and optical waveguide arrays.
In future, it would be interesting to extend our scheme to measure topological invariants in high-dimensional systems, multi-band systems, periodically driven systems and disordered systems.

%%%%%%%%%%%%%%%%%%%%%%%%%%%%%%%%%%%%%%%%%%%%%%%%%

\begin{acknowledgments}
B.Z. and Y.K. made equal contributions. The authors thank Yuri S. Kivshar, Andrey A. Sukhorukov, Zhihuang Luo, Yuangang Deng, Shi Hu, Ling Lin, and Zhoutao Lei for discussions. This work is supported by the National Natural Science Foundation of China (NNSFC) under Grants No. 11374375, No. 11574405, No. 11805283 and No. 11904419, the Hunan Provincial Natural Science Foundation under Grants No. 2019JJ30044, and the International Postdoctoral Exchange Fellowship Program No. 20180052.
\end{acknowledgments}


%\end{document}

%%%%%%%% Supplementary Material %%%%%%%%
\onecolumngrid
\clearpage

\begin{center}
\noindent\textbf{\large{Supplemental Material}}
\onecolumngrid
\end{center}

\setcounter{equation}{0}
\newcounter{sfigure}
\setcounter{sfigure}{1}
\setcounter{table}{0}
\renewcommand{\theequation}{S\arabic{equation}}

 \renewcommand\thefigure{S{\arabic{figure}}}
\renewcommand{\thesection}{S\arabic{section}}
%\renewcommand{\thesubsection}{S\arabic{subsection}}
% \renewcommand{\bibnumfmt}[1]{[S#1]}

%\tableofcontents

%%%%%%%%%%%%%%%%%%%%%%%%%%%%%%%%%%%%%%%%%%
\section{S1. Convergence of dynamic winding number\label{Sec1}}
According to Eq.(2) in the main text, it seems that the definition of dynamic winding number depends on the initial state, and it is unclear whether such number is convergent in the long time.
Here, we prove that the initial state can be rather general and the dynamic winding number is convergent.
The time-average of $\langle\sigma_j\rangle$ is given as
%\begin{widetext}
\begin{eqnarray}
\overline{\sigma_j}=\lim_{T\rightarrow \infty}\frac{1}{T}\int_0^{T}\frac{\sum_{\mu,\mu'}c_{\mu}c_{\mu'}^*e^{-i(\varepsilon_{\mu}-\varepsilon_{\mu'}^*)t}\langle\chi_{\mu'}|\sigma_j|\varphi_{\mu}\rangle}
  {\sum_{\mu}|c_{\mu}|^2e^{-i(\varepsilon_{\mu}-\varepsilon_{\mu}^*)t}}dt.
  \label{osigmajt}
\end{eqnarray}
For Hermitian systems, $\varepsilon_{\mu}=\varepsilon_{\mu}^*$, and the periodic terms vanish in the long time average and only the diagonal terms preserve.
The above equation can be simplified as
\begin{eqnarray}
   \overline{\sigma_j}= \sum\limits_{\mu}|c_{\mu}|^2
  \langle\chi_{\mu}|\sigma_j|\varphi_{\mu}\rangle=\big(|c_{+}|^2-|c_{-}|^2\big)\frac{h_j}{\varepsilon_{+}}
\end{eqnarray}
For the non-Hermitian systems, we assume that the eigenenergy $\varepsilon_\mu=\mu(A+iB)$, where $A$ and $B$ are real numbers.
When $B>0$, Eq.~\eqref{osigmajt} is approximately given as
\begin{eqnarray}
    \overline{\sigma_j}&=&\langle\chi_{+}|\sigma_j|\varphi_{+}\rangle=\frac{h_j}{\varepsilon_{+}}. \label{sigmajplus}
\end{eqnarray}
Similarly, when  $B<0$, Eq.~\eqref{osigmajt} is approximately given as
\begin{eqnarray}
    \overline{\sigma_j}&=&\langle\chi_{-}|\sigma_j|\varphi_{-}\rangle=-\frac{h_j}{\varepsilon_{+}}. \label{sigmajminus}
\end{eqnarray}
Combining with Eqs.~\eqref{sigmajplus} and  \eqref{sigmajminus}, we can also obtain
\begin{eqnarray}
  \frac{\overline{\sigma_j}}{\overline{\sigma_i}}=\frac{h_j}{h_i},
  \label{osigmajnj}
\end{eqnarray}
in the conditions $|c_{+}|^2\ne|c_{-}|^2$ for Hermitian systems and  $c_{+}\ne 0$ for $B>0$ or $c_{-}\ne 0$ for $B<0$ in non-Hermitian systems.
It means that the dynamic winding number also converges in the long time limits when the initial state satisfies a few constraint.
According to the Eq.~\eqref{osigmajnj}, one can also obtain
\begin{eqnarray}
\eta_{ji}=\phi_{ji}=\arctan \left(\frac{\langle\chi_{\mu}|\sigma_j|\varphi_{\mu}\rangle}{\langle\chi_{\mu}|\sigma_i|\varphi_{\mu}\rangle}\right),
\label{osigmajnj_1}
\end{eqnarray}
where the azimuthal angle $\eta_{ji}=\arctan(\overline{\sigma_j}/\overline{\sigma_i})$ and $\phi_{ji}=\arctan(h_j/h_i)$.

\section{S2. Relation between dynamic winding number and time-averaged spin textures \label{Sec2}}
For the non-Hermitian case, the azimuthal angle $\eta_{ji}$ and $\phi_{ji}$ is generally a complex angle, so that they do not represent physical observables in the biorthogonal system.
This problem can be fixed by  decomposing the azimuthal angle into two parts, $\phi_{ji}=\Re(\phi_{ji})+\Im(\phi_{ji})$, where $\Re(\phi_{ji})$ and $\Im(\phi_{ji})$ represents the real part and image part of $\phi_{ji}$.
The azimuthal angle satisfies
\begin{eqnarray}
&&e^{i2\phi_{ji}}=e^{i2\Re(\phi_{ji})}e^{-2\Im(\phi_{ji})}=\frac{1+i\tan(\phi_{ji})}{1-i\tan(\phi_{ji})}=\frac{h_i+ih_j}{h_i-ih_j},
\nonumber \\
&&e^{-2\Im(\phi_{ji})}=\big|\frac{h_i+ih_j}{h_i-ih_j}\big|,
\label{osigmajnj_2}
\end{eqnarray}
$\Re(\phi_{ji})$ and $\Im(\phi_{ji})$ contribute to the argument and amplitude, respectively.
%Excluding EPs which satisfy $h_i^2+h_j^2=0$,
$\Im(\phi_{ji})$ is a real continuous periodic function of $\bm{k}$, so that $\oint_{S} \partial_{\bm k}\Im[\phi_{ji}(\bm k)]d\bm{k}=\oint_{S} \partial_{\bm k}\Im[\eta_{ji}(\bm k)]d\bm{k}=0$. It means that only the real part of azimuthal angle contributes to the dynamic winding number,
\begin{eqnarray}
w_d=\frac{1}{2\pi}\oint_{S} \partial_{\bm k}\Re(\eta_{ji}) d\bm{k}=\frac{1}{2\pi}\oint_{S} \partial_{\bm k}\Re(\phi_{ji}) d\bm{k},
\label{dynwinding_20190628}
\end{eqnarray}
Next, we will show that the real part of azimuthal angle is a physical observable.
According to the Eq.~\eqref{osigmajnj_2}, the real part of azimuthal angle satisfies
\begin{eqnarray}
\tan(2\Re(\phi_{ji}))=\frac{\Im\big(\frac{\pounds h_i+i\pounds h_j}{\pounds h_i-i\pounds h_j}\big)}{\Re\big(\frac{\pounds h_i+i\pounds h_j}{\pounds h_i-i\pounds h_j}\big)},
\end{eqnarray}
where $\pounds$ is a nonzero arbitrary constant.
After some algebras, one can rewrite the above relation as
\begin{eqnarray}
\tan(2\Re(\phi_{ji}))=\frac{\tan(\phi_{ji}^{RR})+\tan(\phi_{ji}^{LL})}{1-\tan(\phi_{ji}^{RR})\tan(\phi_{ji}^{LL})}=\tan(\phi_{ji}^{RR}+\phi_{ji}^{LL}), \nonumber
\end{eqnarray}
where
\begin{eqnarray}
\tan(\phi_{ji}^{RR})=\frac{\Re(\pounds h_j)+\Im(\pounds h_i)}{\Re(\pounds h_i)-\Im(\pounds h_j)}, \nonumber \\
\tan(\phi_{ji}^{LL})=\frac{\Re(\pounds h_j)-\Im(\pounds h_i)}{\Re(\pounds h_i)+\Im(\pounds h_j)},
\label{phi_ab}
\end{eqnarray}
which define two real angles $\phi_{ji}^{RR}$ and $\phi_{ji}^{LL}$, respectively.
It is worth noting that the two real angles $\phi_{ji}^{RR}$ and $\phi_{ji}^{LL}$ will be changed by different parameters $\pounds$, but $\Re(\phi_{ji})$ still keeps the same.
The relation between $\Re(\phi_{ji})$ and $\phi_{ji}^{RR}, \phi_{ji}^{LL}$ satisfies
\begin{eqnarray}
\Re(\phi_{ji})=\Re(\eta_{ji})=\frac{1}{2}(\phi_{ji}^{RR}+ \phi_{ji}^{LL})+n \frac{\pi}{2},
\label{phi_ab17}
\end{eqnarray}
where $n$ is an integer. It means the dynamic winding number $w_d=\frac{1}{2}(w_d^{RR}+w_d^{LL})$.
Here, $w_d^{\tau}=\frac{1}{2\pi}\oint_S \partial_{\bm{k}}\phi_{jl}^{\tau} d\bm{k}$, $\tau\in RR, LL$.
Interestingly, the two real angles $\phi_{ji}^{RR}$ and $\phi_{ji}^{LL}$ can be respectively replaced by time-averaged spin textures corresponding to $|\psi_{\bm{k}}(t)\rangle$ and $|\tilde{\psi}_{\bm{k}}(t)\rangle$,
\begin{eqnarray}
&&\phi_{ji}^{RR}=\arctan\left(\frac{\overline{\langle\psi_{\bm{k}}(t) |\sigma_j|\psi_{\bm{k}}(t)\rangle}}{\overline{\langle\psi_{\bm{k}}(t) |\sigma_i|\psi_{\bm{k}}(t)\rangle}}\right),\nonumber \\
%\frac{\langle\varphi_{\bm{k}+}|\sigma_j|\varphi_{\bm{k}+}\rangle}{\langle\varphi_{\bm{k}+}|\sigma_i|\varphi_{\bm{k}+}\rangle},
&&\phi_{ji}^{LL}=\arctan\left(\frac{\overline{\langle\tilde{\psi}_{\bm{k}}(t) |\sigma_j|\tilde{\psi}_{\bm{k}}(t)\rangle}}{\overline{\langle\tilde{\psi}_{\bm{k}}(t) |\sigma_i|\tilde{\psi}_{\bm{k}}(t)\rangle}}\right),
\label{guanxishi_20190630}
\end{eqnarray}
where $\overline{\langle \bullet \rangle}= \lim_{T \rightarrow \infty} \frac{1}{T}\int_0^{T} \langle \bullet \rangle dt$.
The Eqs.~\eqref{guanxishi_20190630} indicates that the two real angles $\phi_{ji}^{RR}$ and $\phi_{ji}^{LL}$ are physical observables.
For simplicity, we prove the relation with two real angles, $\phi_{yx}^{RR}$ and $\phi_{yx}^{LL}$ in the case of $B>0$.
%
%Case I: $B>0$.
%
The right-right spin textures defined with $|\psi_{\bm{k}}(t)\rangle$ satisfies
\begin{eqnarray}
&&\frac{\overline{\langle\psi_{\bm{k}}(t) |\sigma_y|\psi_{\bm{k}}(t)\rangle}}{\overline{\langle\psi_{\bm{k}}(t) |\sigma_x|\psi_{\bm{k}}(t)\rangle}}=\frac{\langle\varphi_{+}|\sigma_y|\varphi_{+}\rangle}{\langle\varphi_{+}|\sigma_x|\varphi_{+}\rangle},
\label{spin polarization_20190630}
\end{eqnarray}
and the left-left spin textures defined with $|\tilde{\psi}_{\bm{k}}(t)\rangle$ satisfies
\begin{eqnarray}
\frac{\overline{\langle\tilde{\psi}_{\bm{k}}(t)|\sigma_y|\tilde{\psi}_{\bm{k}}(t)\rangle}}{\overline{\langle\tilde{\psi}_{\bm{k}}(t) |\sigma_x|\tilde{\psi}_{\bm{k}}(t)\rangle}}=\frac{\langle\chi_{+}|\sigma_y|\chi_{+}\rangle}{\langle\chi_{+}|\sigma_x|\chi_{+}\rangle}.
\label{spin polarization_20190631}
\end{eqnarray}
According to the Hamiltonian (1) in main text, neither the eigenstates $|\varphi_{\mu}\rangle$ nor $\langle\chi_{\mu}|$ are orthogonal in the non-Hermitian system.
We adopt biorthogonal vectors which fulfill $\langle\chi_{\nu}| \varphi_{\mu}\rangle=\delta_{\nu,\mu}$ and $\sum_{{\mu}}|\varphi_{\mu} \rangle \langle \chi_{\mu} |=1$ by normalizing  $|\varphi_{\mu}\rangle=|\varphi_{\mu}\rangle/N_{\mu}$ and $\langle\chi_{\mu}|=\langle\chi_{\mu}|/N_{\mu}$ with $N_{\mu}=\sqrt{\langle\chi_{\mu}| \varphi_{\mu}\rangle}$, this is,
\begin{eqnarray} \label{equaion1a}
|\varphi_{\mu}\rangle&=&\frac{1}{\sqrt{2\varepsilon_{\mu}(\varepsilon_{\mu}-h_z)}}(h_x-ih_y,\varepsilon_{\mu}-h_z)^{\hat{T}}, \nonumber \\
\langle \chi_{\mu}|&=&\frac{1}{\sqrt{2\varepsilon_{\mu}(\varepsilon_{\mu}-h_z)}}(h_x+ih_y,\varepsilon_{\mu}-h_z),
\label{BZT}
\end{eqnarray}
where the superscript $\hat{T}$ is the transpose operation.
Combining with Eq.~\eqref{spin polarization_20190630}, \eqref{spin polarization_20190631} and \eqref{BZT}, we can immediately obtain,
\begin{eqnarray}
&&\frac{\langle\varphi_{+}|\sigma_y|\varphi_{+}\rangle}{\langle\varphi_{+}|\sigma_x|\varphi_{+}\rangle}= \frac{\Re(h_y\pounds_1)+\Im(h_x\pounds_1)}{\Re(h_x\pounds_1)-\Im(h_y\pounds_1)}, \nonumber \\
&&\frac{\langle\chi_{+}|\sigma_y|\chi_{+}\rangle}{\langle\chi_{+}|\sigma_x|\chi_{+}\rangle}= \frac{\Re(h_y\pounds_1)-\Im(h_x\pounds_1)}{\Re(h_x\pounds_1)+\Im(h_y\pounds_1)},
\label{zhengming_20190630}
\end{eqnarray}
where $\pounds_1=h_z^*+\varepsilon_{+}^*$. Similarly, one can obtain $\pounds_1=h_z^*-\varepsilon_{+}^*$ for the case of $B<0$.
Combining with Eq.~\eqref{phi_ab} and \eqref{zhengming_20190630}, one can easily obtain the relations of Eq.~\eqref{guanxishi_20190630}.

\section{S3. Dynamic winding number in the presence/absence of chiral symmetry \label{Sec3}}
Winding number has been widely used for characterizing the topology of Hermitian systems with chiral symmetry.
In one dimension, winding number can be applied to both Hermitian and non-Hermitian systems with or without chiral symmetry.
Here, we consider a 1D two-band topological system governed by the Hamiltonian,
\begin{eqnarray}
  H(k)=h_x(k)\sigma_x+h_y(k)\sigma_y+h_z(k)\sigma_z. \label{equaion1aa}
\end{eqnarray}
The conventional winding number for each band is defined as~\cite{syin2018geometrical, sjiang2018topological},
\begin{eqnarray}
w_\mu=\frac{1}{\pi}\oint_S dk \langle\chi_\mu|i\partial_k|\varphi_\mu\rangle
=\frac{1}{2\pi}\oint_S dk \frac{h_x\partial_k h_y-h_y\partial_k h_x}{\varepsilon_{\mu}(\varepsilon_{\mu}-h_z)}.
\end{eqnarray}
where $S$ is a closed loop with $k$ varying from $0$ to $2\pi$.
Next, we will build relation between conventional winding number to the dynamic winding number in different situations.

\subsection{A. Chiral symmetric systems} \label{ChiralSystem}
When $h_z=0$, the Hamiltonian~\eqref{equaion1aa} has chiral symmetry $\Gamma H(k) \Gamma =-H(k)$ with $\Gamma=i \sigma_x \sigma_y$.
The conventional winding numbers for different bands are the same, and we denote as
\begin{eqnarray} \label{equaion1b}
w_{\pm }=\frac{1}{2\pi}\oint_S dk \frac{h_x\partial_k h_y-h_y\partial_k h_x}{h_x^2+h_y^2}.
\end{eqnarray}
The expression reduces to the Hermitian cases when $\langle\chi_{\mu}|=\langle\varphi_{\mu}|$.
If we define an azimuthal angle as $\phi_{yx}=\arctan(h_y/h_x)$, the above equation is given as
\begin{eqnarray}
 w_{\pm}=\frac{1}{2\pi}\oint_S \partial_k \phi_{yx} dk,
\end{eqnarray}
According to Eq.~\eqref{dynwinding_20190628}, we can immediately conclude that the conventional winding number is equal to the dynamic winding number,
\begin{eqnarray}
w_{\pm}=w_d,
\end{eqnarray}
under  a few constraints of initial state:  $|c_{+}|^2\ne|c_{-}|^2$ for Hermitian systems and $|c_{+}|^2\ne 0\bigcap |c_{-}|^2\ne 0$ for non-Hermitian systems.

\begin{figure}[htp]
\center
\includegraphics[width=4in]{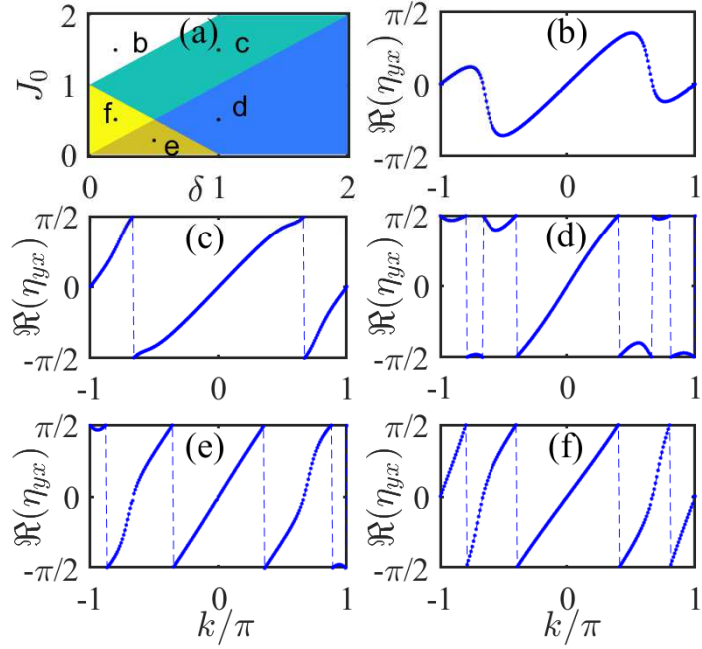}
  \caption{(a) The phase diagram of the 1D chiral-symmetric topological systems.
  The white, blue, green, dark-yellow and bright-yellow regions respectively share winding number as $w_{\pm}=0, 1/2, 1, 3/2$, and $2$.
  (b)-(f) $\Re(\eta_{y,x})$ as a function of $k$ in different parameters ($J_0,\delta$), which are ($1.5,0.2$), ($1.5,1$), ($0.5,1$), ($0.2,0.5$) and ($0.5,0.2$), corresponding to points $b, c, d, e$ and $f$ in (a), respectively. The other parameters are chosen as $J_1=1$, $J_2=1$ and  $h_z=0$.}\label{fig2b}
\end{figure}

To numerically verify our theory, we consider the systems with $h_x=J_0+J_1\cos(k)+J_2\cos(2k)$ and $h_y=J_1\sin(k)+J_2\sin(2k)-i\delta$.
In the non-Hermitian case with $\delta\neq0$, the conventional winding numbers $w_{\pm}$ can appear half integer values in some parameter ranges, different from the integer values in the Hermitian systems.
For simplicity, we take $J_1=1$, and $J_0$, $J_2$, $\delta$ are real.
The dispersion of this Hamiltonian is
\begin{eqnarray} \label{equaion10b}
\varepsilon_{\pm}(k)=\pm\sqrt{(J_0-\delta+e^{-ik}+J_2e^{-2ik})(J_0+\delta+e^{ik}+J_2e^{2ik})}.
\end{eqnarray}
The energy is symmetric about zero energy, which is ensured by the chiral symmetry.
Since the energy gap must close at phase transition points, we can determine the phase boundaries by the band-crossing condition $\varepsilon_{\pm}(k)=0$, which yields $J_0=\pm \delta+1-J_2$ and $J_0=\pm \delta-1-J_2$ for arbitrary $J_2$. Particularly, $J_0=J_2\pm\delta$ if $|J_2|>0.5$.
Fixing $J_1=1$, $J_2=1$ and $h_z=0$ and changing both $\delta$ and $J_0$, we calculate topological phase diagram distinguished by their winding numbers, see Fig.~\ref{fig2b} (a).
Here, the white, blue, green, dark-yellow and bright-yellow regions possess conventional winding number $w_{\pm}=0, 1/2, 1, 3/2$ and $2$, respectively.
In Figs.~\ref{fig2b} (b)-(f), we also give the angle $\Re(\eta_{yx})$ as a function of quasi-momentum $k$ with different parameters ($J_0,\delta$) marked as b, c, d, e, f in the Figs.~\ref{fig2b} (a).
 The dynamic winding number are $0$, $1$, $1/2$, $3/2$ and $2$, respectively.
The numerical results are in well agreement with the theoretical prediction, which prove the validity for our dynamic approach once again.

\subsection{B. Non-chiral symmetric systems}
\begin{figure}[htp]
\center
\includegraphics[width=4in]{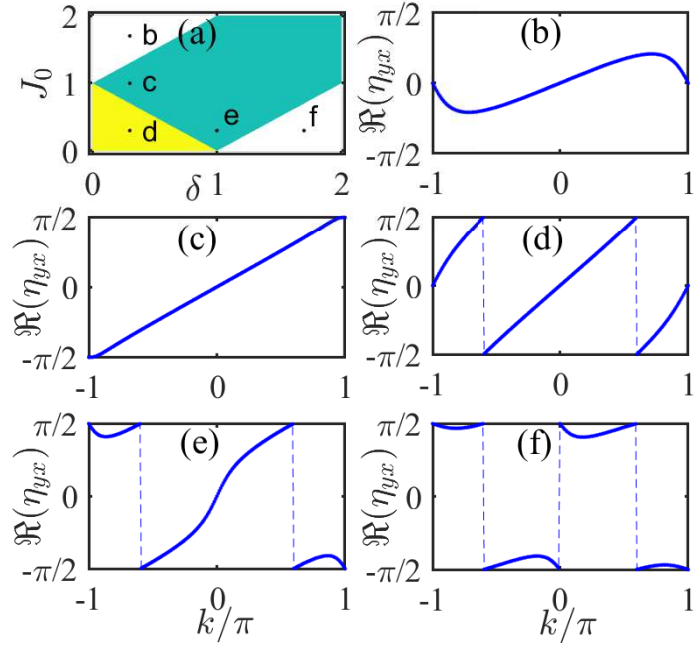}
  \caption{(a) The phase diagram of the 1D non-chiral-symmetric topological systems.
  The white, green and yellow regions share winding number as $w_{t}=0, 1$,and $2$, respectively.
  (b)-(f) $\Re(\eta_{yx})$ as a function of $k$ in different parameters ($J_0,\delta$), which are ($1.7,0.3$), ($1,0.3$), ($0.3,0.3$), ($0.3,1$) and ($0.3,1.7$), corresponding to points $b, c, d, e$ and $f$ in (a), respectively. The other parameters are chosen as $J_1=1$, $J_2=0$ and $h_z=0.5$.}\label{fig1_nochiral_2019}
\end{figure}

When $h_z\ne 0$, the Hamiltonian~\eqref{equaion1aa} breaks the chiral symmetry.
Unlike the systems with chiral symmetry, the conventional winding number for each band is not a quantized number, which indicates that $w_{\pm}$ is no longer a topological invariant.
However, the sum of the winding numbers for different bands,
\begin{eqnarray} \label{equaion1d}
w_{t}=w_{+}+w_{-}=\frac{1}{\pi}\oint_S dk \frac{h_x\partial_k h_y-h_y\partial_k h_x}{h_x^2+h_y^2},
\end{eqnarray}
has been demonstrated to be a topological invariant~\cite{sjiang2018topological}.
%
%When $h_z\ne 0$, the Hamiltonian~\eqref{equaion1aa} breaks the chiral symmetry.
%%
%The summation of the winding number of different bands, $w_{t}=\omega_++\omega_-$, has been demonstrated to be a topological invariant\cite{jiang2018topological}.
%So that one can obtain the new topological invariant
%\begin{eqnarray} \label{equaion1d}
%w_{t}=\omega_++\omega_-=\frac{1}{\pi}\oint_c dk \frac{h_x\partial_k h_y-h_y\partial_k h_x}{h_x^2+h_y^2}.
%\end{eqnarray}
The topological invariant $w_{t}$ is independent of $h_z$, although its definition is related to the eigenvector of $H(k)$.
The parameters $h_x$ and $h_y$ become very important for the definition of topological invariant.
Except for the exceptional point $h_x^2+h_y^2=0$, we introduce a complex angle $\phi_{yx}$ satisfying $\tan(\phi_{yx})=h_y/h_x$.
In terms of $\phi_{yx}$, $w_{t}$ can be represented as
\begin{eqnarray} \label{equaion1f}
&&w_{t}=\frac{1}{\pi}\oint_S \partial_k\phi_{yx} dk,
\end{eqnarray}
where the integral is also taken along a loop with $k$ from $0$ to $2\pi$.
According to Eq.~\eqref{sdynwinding_20190628}, we can relate the topological invariant $w_{t}$ to the dynamic winding number
\begin{eqnarray}
  w_{t}=w_++w_-= 2 w_d,
\end{eqnarray}
under a few constraints of initial state:  $|c_{+}|^2\ne|c_{-}|^2$ for Hermitian systems and $|c_{+}|^2\ne 0\bigcap |c_{-}|^2\ne 0$ for non-Hermitian systems.
Fixing $J_1=1$, $J_2=0$ and $h_z=0.5$ in the same model as that in Subsec.~\ref{sChiralSystem}, we calculate the topological invariant $w_{t}$ as a function of $\delta$ and $J_0$, see Fig.~\ref{fig1_nochiral_2019} (a).
Here, the white, green, and bright-yellow regions possess topological invariant $w_{t}=0, 1$ and $2$, respectively.
In Figs.~\ref{fig1_nochiral_2019} (b)-(f), we also give the angle $\Re(\eta_{yx})$ versus the quasi-momentum $k$ with different parameters ($J_0,\delta$) marked as b, c, d, e, f in the Fig.~\ref{fig1_nochiral_2019} (a).
The dynamic winding number are $0$, $1/2$, $1$, $1/2$ and $0$, respectively.
The numerical results are also in well agreement with the theoretical prediction, which demonstrate the validity of our dynamic approach.

\section{S4. Chern number in 2D systems  \label{Sec4}}
\subsection{A. Alternate choice of reference axis}
\begin{figure}[htp]
\center
\includegraphics[width=4.5in]{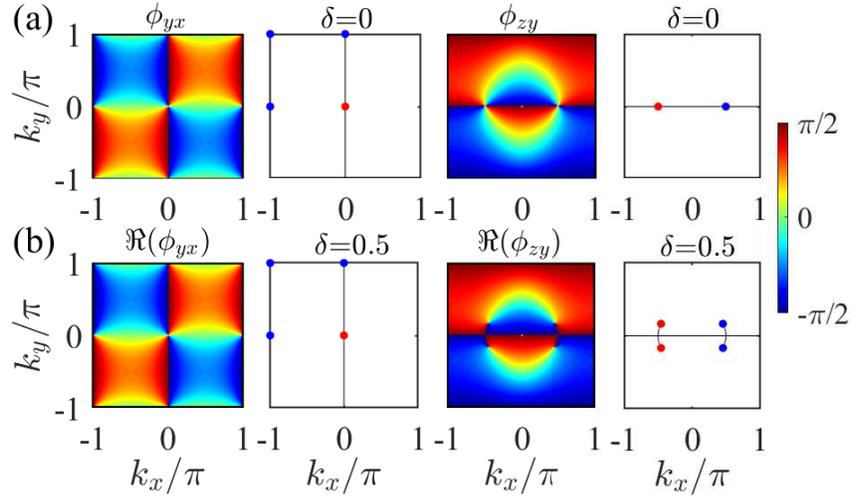}
  \caption{Topologically nontrivial phase with Chern number $C=1$.
  Hermitian case: (a) Azimuthal angle $\phi_{yx}$ and $\phi_{zy}$ in parameter space ($k_x,k_y$),
  where blue and red points represent north and south poles of the Bloch spherical surface.
  Non-Hermitian case: (b) azimuthal angle $\Re(\phi_{yx})$ and $\Re(\phi_{zy})$ in parameter space ($k_x,k_y$).
  where blue and red points represent north and south EPs of the virtual Bloch spherical surface.
}\label{fig444}
\end{figure}
In our main text, we only consider $\theta=\theta_y=\arccos(h_y/|\vec{h}(\bm{k})|)$ and $\phi=\phi_{xz}=\arctan(h_x/h_z)$.
Alternatively, we can also take $\theta=\theta_z=\arccos(h_z/|\vec{h}(\bm{k})|)$  and $\phi=\phi_{yx}=\arctan(h_y/h_x)$, or $\theta=\theta_x=\arccos(h_x/|\vec{h}(\bm{k})|)$  and $\phi=\phi_{zy}=\arctan(h_z/h_y)$.
These two choices lead to distinct observations, but give the same Chern number.
In Fig.~\ref{fig444}(a) and (b), based on different choices of reference axis, we give the azimuthal angle $\phi_{yx}$ and $\phi_{zy}$ in Hermitian($\delta=0$) and non-Hermitain($\delta=0.5$) cases, where the other parameters are the same as the Fig. 2(c) of the main text.
Around the  singularity points, one can also easily obtain the dynamic winding number, and the Chern number $C=1$ in both the Hermitian and non-Hermitian cases, consistent with the ideal Chern number.
This is because the different references only differ from a gauge transformation and the Chern number do not depend on the choice of reference.

\subsection{B. Larger Chern number  \label{Sec5}}

\begin{figure}[!htp]
\center
\includegraphics[width=4in]{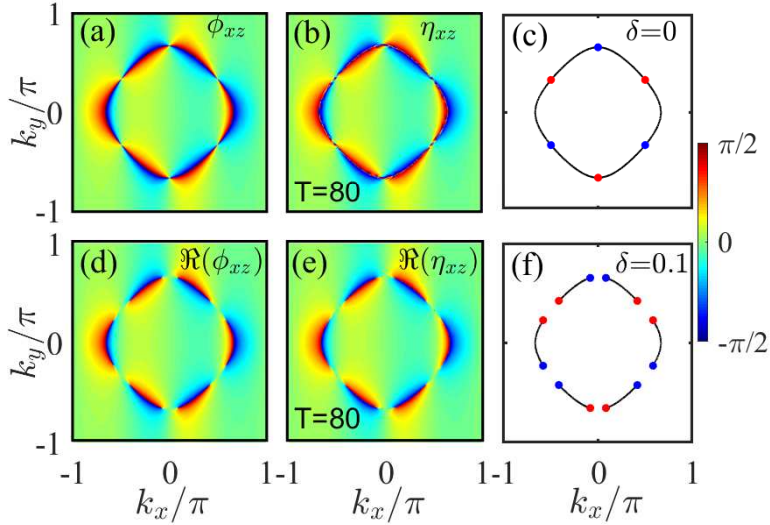}
  \caption{Topologically nontrivial phase with Chern number $C=3$.
  Hermitian case at top: (a) and (b) correspond to the azimuthal angle $\phi_{xz}$ and $\eta_{xz}$ in parameter space ($k_x,k_y$).
  (c) Blue and red points represent north and south poles of the Bloch spherical surface.
  Non-Hermitian case at bottom: (d) and (e) respectively correspond to the azimuthal angle $\Re(\phi_{xz})$ and $\Re(\eta_{xz})$ in parameter space ($k_x,k_y$).
  (f) Blue and red points represent north and south EPs of the  virtual Bloch spherical surface.
}\label{fig5s8}
\end{figure}
The dynamic approach is clearly applicable to topological phases with larger Chern numbers.
To show this, we consider another two-band model which supports band structure with larger Chern number, this is,
\begin{eqnarray} \label{equaion10sa66}
h_x&=&J_x \sin(2k_x); h_y=J_y \sin(2k_y); h_z=m_z-J_z \cos(k_x)-J_z \cos(k_y)-i \delta.
\end{eqnarray}
For the Hermitian case $\delta=0$, the trivial phase is lying in $|m_z|>2J_z$, while the topological phases are distinguished as: ($i$) $J_z<m_z<2J_z$ with the Chern number $C=-1$; ($ii$) $0<m_z<J_z$ with $C=3$; ($iii$) $-J_z<m_z<0$ with $C=-3$; ($iv$) $-2J_z<m_z<-J_z$ with $C=1$.
Here we only verify topological phase with Chern number $C=3$, where the other parameters are chosen as $J_x=J_y=0.2,\ J_z=1$ and $m_z=0.5$.
In Figs.~\ref{fig5s8}(a) and \ref{fig5s8}(b), we give the azimuthal angle $\phi_{xz}$ and $\eta_{xz}$ in parameter space $(k_x,k_y)$, respectively.
One can find that more north and south poles appear in Fig.~\ref{fig5s8}(c), compared with Fig. 2 in the main text.
From the Eq.~(10) in main text, one can also easily obtain the Chern number $C=3$ via dynamic winding number.
For the non-Hermitian case, we consider $\delta=0.1$ and the other parameters are the same as those in the Hermitian case.
%
%This non-Hermitian system is topological and the Chern number $C =3$.
In Figs.~\ref{fig5s8}(d) and \ref{fig5s8}(e), we also give the azimuthal angle $\Re(\phi_{xz})$ and $\Re(\eta_{xz})$ in parameter space $(k_x,k_y)$, respectively.
The dynamic winding numbers around EPs become half, while the EPs become double as the Hermitian counterpart, see Fig.~\ref{fig5s8}(f).
Eventually, the Chern number keeps the same as that in the Hermitian case.

\section{S5. Experimental consideration \label{Sec6}}
One can immediately apply the dynamical approach for topological Hermitian systems.
Cold atom systems is an excellent platform to realize topological band models and detect topological invariants.
One and two dimensional spin-orbit couplings have been realized in a highly controllable Raman lattice \cite{squ2013observation, shamner2014dicke, ssun2018highly, ssun2018uncover}.
Initial states are quite easily prepared by loading the atoms into the lattices.
Here, the initial constraint $|c_{\bm{k}+}|^2\ne |c_{\bm{k}-}|^2$ may be not satisfied for some specific momentum $\bm k$, but the occurred probability is so small that the global dynamical azimuthal angle is not affected due to the topological nature.
The spin population $N_{\uparrow(\downarrow)}(\bm{k})$ with different momentum can be measured by spin-resolved time-of-flight (TOF)  absorption imaging\cite{ssun2018uncover}.
Thus, one can obtain the spin population difference $\langle\psi_{\bm{k}}(t)| \sigma_z | \psi_{\bm{k}}(t)\rangle=(N_{\uparrow}(\bm{k})-N_{\uparrow}(\bm{k}))/(N_{\uparrow}(\bm{k})+N_{\uparrow}(\bm{k}))$.
The spin textures $\langle\psi_{\bm{k}}(t)| \sigma_{x(y)} | \psi_{\bm{k}}(t)\rangle$ can be transferred to the spin population difference by applying $\pi/2$ pulse,
that is, $\langle\psi_{\bm{k}}(t)| \sigma_{x(y)} | \psi_{\bm{k}}(t)\rangle=\langle\psi_{\bm{k}}(t)|e^{-i\frac{\pi}{2}\frac{\sigma_{y(x)}}{2}} \sigma_{z}e^{i\frac{\pi}{2}\frac{\sigma_{y(x)}}{2}}| \psi_{\bm{k}}(t)\rangle$.
Because the cold atom systems have long coherent time, there is no obstacle to extract the dynamic winding number via long time average of the spin textures.

To apply the dynamical approach in topological non-Hermitian systems, we should first consider how to realize the topological non-Hermitian models in experiments.
Since two-level non-Hermitian models have been widely realized in optical systems, such as two coupled optical cavities~\cite{schang2014parity, sPeng2014}, optical waveguides~\cite{sgordon2000pmd, sruter2010observation, szeuner2015observation}, optomechanical cavity\cite{saspelmeyer2014cavity, sxu2016topological, sverhagen2017optomechanical} etc.
We mainly discuss how to extract dynamic winding number with two optical waveguides with tunable parameters.
A two-level non-Hermitian system can be realized by introducing gain and loss in the two waveguides.
The coupling strength can be tuned by the waveguide separation.
We regard the two different waveguides as two spin components.
The initial states can be prepared by randomly split the light injecting into the two waveguides.
One can obtain $\langle\psi(l)|\sigma_z|\psi(l)\rangle$ by measuring the intensity difference between two waveguides at propagating distance $l$. Here, the distance $l$ plays the role of time.
Actually, the final states will collapse into one of the eigenstate in the long distance.
Thus, the output intensity difference of the waveguides is sufficient and long distance average of the intensity difference is not necessary.
One can also obtain  $\langle\psi(l)|\sigma_{x(y)}|\psi(l)\rangle$ by insetting a beam splitter before intensity measurement.
By designing the waveguide separation and gain and loss rates, one can simulate the two-band model.
Repeating the above operations, one can finally construct the dynamic winding number.

\end{document}